\documentclass[12pt]{iopart}
\usepackage{epsfig}
\usepackage{color}
\newcommand{\gtrsim}{\,\rlap{\lower4pt\hbox{$\mathchar\sim$}}
\raise1.8pt\hbox{$>$}\,}
\newcommand{\lesssim}{\,\rlap{\lower4pt\hbox{$\mathchar\sim$}}
\raise1.8pt\hbox{$<$}\,}

\begin{document}
						
\title[Supernova neutrino three-flavor evolution  
            with dominant collective effects]{Supernova neutrino three-flavor evolution
            with dominant collective effects}
   
\author{Gianluigi Fogli$^{1,2}$, 
		Eligio Lisi$^2$, 
		Antonio Marrone$^{1,2}$ and 
 		\\ Irene Tamborra$^{1,2}$ }
\address{$^1$~Dipartimento Interateneo di Fisica ``Michelangelo Merlin,'' \\
		Via Amendola 173, 70126 Bari, Italy} 
%%%%%%% ++++++++++
\address{$^2$~Istituto Nazionale di Fisica Nucleare, Sezione di Bari,\\
         Via Orabona 4, 70126 Bari, Italy}
%%%%%%% ++++++++++

%\date{April 2009}

\begin{abstract}
Neutrino and antineutrino fluxes from a 
core-collapse galactic supernova are studied, within a representative three-flavor
scenario with inverted mass hierarchy and tiny 1-3 mixing. 
The initial flavor evolution is dominated by collective self-interaction effects,
which are computed in a full three-family framework along an averaged radial
trajectory. During the whole time span considered ($t=1$--20~s), 
neutrino and antineutrino spectral splits emerge as dominant features in the
energy domain for the final, observable fluxes. The main results can be useful for SN event rate
simulations in specific detectors. 
Some minor or unobservable three-family features (e.g, related to the muonic-tauonic flavor 
sector), as well as observable effects due to variations in the spectral input,
are also discussed for completeness. 
\end{abstract}

\pacs{14.60.Pq, 13.15.+g, 97.60.Bw}

\maketitle

%%%%%%%%%%%%%%%%%%%%%%%%%%%%%%%%%%%%%%%%%%%%%%%%%%%%%%%%%%%%%%%%%%%%%%
\section{Introduction} %%%%%%%%%%%%%%%%%%%%%%%%%%%%%%%%%%%%%%%%%%%%%%%
%%%%%%%%%%%%%%%%%%%%%%%%%%%%%%%%%%%%%%%%%%%%%%%%%%%%%%%%%%%%%%%%%%%%%%

The fluxes of $\nu$ and $\overline\nu$ from a core-collapse supernova (SN) can
encode very interesting information about both the (anti)neutrino properties 
\cite{Di08} and the
SN explosion mechanism \cite{Expl}.  Assuming that the latter is understood, one may  try
to detect distinctive features of flavor change in either the energy ($E$) or time
($t$) spectra of the observable fluxes $F_\alpha (E,\,t)$, 
as compared with the ``featureless,'' unoscillated fluxes $F_\alpha^0(E,\,t)$,
for one or more neutrino species $\nu_\alpha$. 

After the recent, seminal work in \cite{Coll,Semi}, it has been fully realized that
flavor change phenomena 
driven by (anti)neutrino self-interactions \cite{Pant} may induce dramatic, observable effects in a variety
of SN scenarios, especially for inverted mass hierarchy \cite{Pend}; see
\cite{Di08,Nonl} for reviews of this rapidly growing research field. In particular, the so-called 
spectral split \cite{Split} or swap \cite{Swap} emerges as a distinctive feature in the neutrino 
\cite{Miri} and possibly the antineutrino \cite{Miri,Tamb} energy spectra. If this
feature dominates, then a  stepwise flavor conversion of 
$\nu_e$ develops across a certain critical energy $E_c$ \cite{Analysis}, namely,
%...............................
\begin{equation}
\label{P}
P_{ee}(E)\simeq \left\{
\begin{array}{ll}
1 & (\mathrm{for\ }E<E_c)\ ,\\
0 & (\mathrm{for\ }E>E_c)\ ,
\end{array}
\right.
\end{equation}
%................................
in terms of the survival probability $P_{ee}=P({\nu_e\to\nu_e})$ 
at the end of collective effects. 
A somewhat similar feature
has been observed in the $\overline\nu_e$ sector \cite{Miri,Tamb}, although 
with at different (generally smaller) split energy $\overline E_c$,
%...............................
\begin{equation}
 \overline P_{ee}(E)\simeq \left\{
\begin{array}{ll}
1 & (E<\overline E_c)\ ,\\
0 & (E>\overline E_c)\ .
\end{array}
\right.
\label{Pbar}
\end{equation}
%................................
In a strictly $2\nu$ framework with flavors ``$e$'' and ``$x$'', 
the net result of collective effects would be a complete interchange, or ``swap,''
of the fluxes $F_\alpha$ between ($\nu_e,\,\nu_x$) above $E_c$
 \cite{Split,Swap,Miri},
as well as between  
$(\overline\nu_e,\,\overline\nu_x)$  above $\overline E_{c}$ 
\cite{Tamb}.

Analogously, in a three-flavor scenario with both $\nu_\mu$ and $\nu_\tau$  acting
as a flavor ``$x$,''
%.........................
\begin{eqnarray}
2F_x 			&=& F_\mu + F_\tau\ ,\\
2\overline F_x 	&=& \overline F_\mu + \overline F_\tau\ ,
\end{eqnarray}
%............................  
and with usual initial conditions 
%......................
\begin{equation}
F^0_\mu=F^0_\tau= F^0_x= \overline F^0_x=\overline F^0_\tau=\overline F^0_\mu\ ,
\end{equation}
%.................................
a swap is expected to occur between the $e$ flavor and one of the two $x$ flavors, namely,
%.............................................................................................
\begin{eqnarray}
\label{je}
F'_e &=& F_e^0 P_{ee} + F^0_x(1-P_{ee}) \simeq \left\{\begin{array}{ll}F^0_{e} & (E<E_c)\ ,\\ F^0_x & (E>E_c)\ ,\end{array} \right.\\
\label{jx}
\overline F'_{ e} &=& \overline F_e^0 \overline P_{ee} + \overline 
F^0_x(1-\overline P_{ee})\simeq  \left\{\begin{array}{ll}\overline F^0_{e} & (E<\overline E_c)\ ,
\\ \overline F^0_{ x} & (E>\overline E_c)\ ,\end{array} \right.
\label{jebar}\\
2F'_x &=& [F_e^0(1-P_{ee})+F_x^0P_{ee}]+F_x^0 \simeq  \left\{\begin{array}{ll}2 F^0_{x} & (E<E_c)\ ,\\ F^0_e + F^0_x & (E>E_c)\ ,\end{array} \right.\\
2\overline  F'_{x}  &=& [\overline F_e^0(1-\overline P_{ee})+\overline F_x^0\overline P_{ee}]+\overline F_x^0 
\simeq  \left\{\begin{array}{ll} 2 \overline F^0_{x} & (E<\overline E_c)\ ,
\\ \overline F^0_{e} + \overline F^0_{ x} & (E>\overline E_c)\ .\end{array} \right.
\label{jxbar}
\end{eqnarray}
%..............................
In the above equations, the primes denote fluxes at the end of collective effects within the SN,
to be further evolved up to the exit from the SN and to the final detector.

It is important to test the above $3\nu$ expectations within comprehensive three-flavor 
calculations, for various reasons. First, one has to show that the
assumed dominance of spectral split phenomena can indeed be manifest (without
being disrupted by ordinary matter effects) in a sufficiently general, uncontrived and interesting 
supernova $3\nu$ scenario.  
Second, the effective
factorization of one family out of  three {\em self-interacting families\/} (leading
to nonlinear equations) is never totally obvious, 
and ``effective $2\nu$'' expectations like
those in Eqs.~(\ref{je})--(\ref{jxbar})
need to be explicitly checked. Third, in the self-interaction context, the
probabilities $P_{\alpha\beta}$ depend, among other things,
on the initial conditions and on the absolute fluxes---which do change if the 
SN energy luminosity (which also varies in time) is distributed over $3\nu$ rather than $2\nu$.
Finally, while the $\nu$ spectral split phenomenon
is robust and largely understood in terms of lepton number conservation
and adiabatic flavor evolution \cite{Simple,Smirnov}, the $\overline \nu$ split
seems to be more fragile and related to (not completely understood) nonadiabatic aspects
of the evolution \cite{Tamb}; it is thus worth checking its appearance in a full $3\nu$ calculation. 

Several recent works have focussed on collective effects in $3\nu$ scenarios 
\cite{Neutroniz,Esteban,Das3,Full3,Basu,Earth,SNDB,Gava} 
and have successfully recovered 
spectral split features in the neutrino sector (and perhaps
also in the antineutrino sector \cite{Das3,Gava}) in inverted
hierarchy. We think it useful to add an independent contribution to this very
recent research field, by discussing a SN $3\nu$ scenario where both the $\nu$ and $\overline\nu$ split features 
are shown to emerge clearly for a relatively long time of $\sim\! 20$ seconds after SN explosion. Such a
scenario is thus particularly suited to prospective experimental tests in high-statistics,
time-integrated energy spectra of events from the next galactic core-collapse SN.

Our work is structured as follows. In Sec.~2 we describe a representative
SN neutrino framework,
where collective effects (in the form of spectral splits) are expected to dominate over ordinary matter effects. 
In Sec.~3 we discuss the $3\nu$ formalism used to compute the flavor evolution of the initial fluxes $F^0_\alpha$.
In Sec.~4 we present our results for the fluxes $F'_\alpha$ at the end of collective effects, 
and show that they confirm the simple expectations in 
Eqs.~(\ref{je})--(\ref{jxbar}); we also discuss minor or unobservable features of our calculations.
In Sec.~5 we complete the flavor evolution of the  observable fluxes $F_\alpha$,
by accounting for final  phase-averaging and matter effects.  In Sec.~6 we discuss the
effects of some variations in the input SN neutrino spectra, with respect to our default scenario.
Section~7 summarizes our results. 
 
The reader is reminded that our results, as well as many other observable features discussed
in the growing literature of self-interaction neutrino effects in supernovae, must be taken with 
a grain of salt. Indeed, many open questions are still open from the viewpoint of
the theory (validity of the mean-field approximation in the neutrino evolution equations), of the 
SN explosion energetics (neutrino luminosities and energy spectra) and geometry (asymmetries, turbulence),
and of the neutrino evolution numerics (convergence and robustness of calculations). Results which
currently appear to be rather generic (such as the spectral split phenomena in inverse hierarchy)
might be unpredictably challenged by improvements and further understanding in any of the above
issues. Hopefully, the next galactic supernova explosion(s) will help to reduce the current level
of uncertainty in the physics and astrophysics hidden in the expected neutrino signal.

%%%%%%%%%%%%%%%%%%%%%%%%%%%%%%%%%%%%%%%%%%%%%%%%%%%%%%%%%%%%%%%%%% 
\section{Reference neutrino parameters and SN scenario}

The adopted $3\nu$ oscillation parameters and our reference SN model are discussed below.

\subsection{Neutrino parameters}

We assume that the hierarchy of neutrino masses $m_i$ is inverted ($m_3<m_{1,2}$), 
%...........................
\begin{equation}
(m^2_1,\,m^2_2,\,m^2_3)=\frac{m^2_1+m^2_2}{2}+\left(-\frac{\delta m^2}{2},\,+\frac{\delta m^2}{2},-\Delta m^2\right)\ ,
\end{equation}
%............................
with squared mass splittings set at the representative values
%...........................
\begin{eqnarray}
\Delta m^2 &=& 2\times 10^{-3}\mathrm{\ eV}^2\ ,\\
\delta m^2 &=& 8\times 10^{-5}\mathrm{\ eV}^2\ .
\end{eqnarray}
%............................. 
The associated ``high'' ($H$) and ``low'' ($L$) vacuum oscillation frequencies are then
%........................................................
\begin{eqnarray}
\omega_H &=& \frac{\Delta m^2}{2E}\label{omegaH} = \frac{5.07}{E/\mathrm{MeV}}\; [\mathrm{km}^{-1}] \ ,\\
\omega_L &=& \frac{\delta m^2}{2E}\label{omegaL} = \frac{0.20}{E/\mathrm{MeV}}\; [\mathrm{km}^{-1}] \ .
\end{eqnarray}
%.........................................................

Within the usual parametrization \cite{PDG} for the mixing matrix 
$U=U(\theta_{12},\,\theta_{13},\,\theta_{23},\,\delta_\mathrm{CP})$,
we fix $\theta_{12}$ as \cite{Latest}
%...................................................
\begin{equation}
\label{theta12}
\sin^2\theta_{12}=0.314\ ,
\end{equation}
%....................................................
while we consider three representative values for the more uncertain 
angle $\theta_{23}$, corresponding to maximal and nonmaximal (but
octant-symmetric) mixing:
%...................................................
\begin{equation}
\label{theta23}
\sin^2\theta_{23}=0.50,\,0.36,\,0.64\ .
\end{equation}
%....................................................
Concerning the third angle $\theta_{13}$, we assume a tiny reference value,
%....................................
\begin{equation}
\label{theta13}
\sin^2\theta_{13}= 10^{-6}\ ,
\end{equation}
%........................................
in order to suppress the impact of ordinary matter effects, as explained in the following Subsection~2.2.
%..................................
Finally, we ignore possible CP-violating effects (which are
arguably very small \cite{Gava}) by setting  
%...................................................
\begin{equation}
\delta_\mathrm{CP}=0\ .
\end{equation}
%....................................................

%%%%%%%%%%%%%%%%%%%%%%%%%%% FIGURE 1 %%%%%%%%%%%%%%%%%%%%%%%%%%%%%%%%
\begin{figure}[t]
\centering
\vspace*{4mm}
\hspace*{22mm}
\epsfig{figure=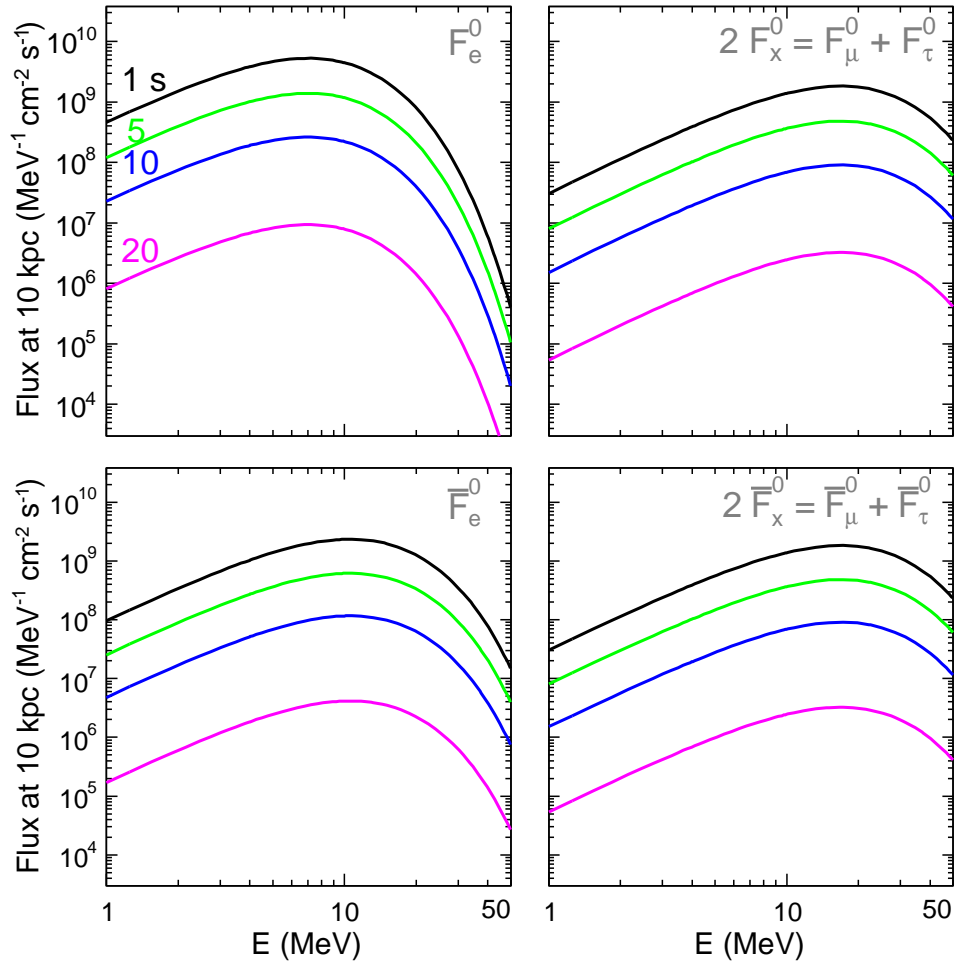,width=0.8\columnwidth}
\vspace*{0mm}
\caption{Unoscillated fluxes of neutrinos
($F^0_\alpha$, top) and antineutrinos ($\overline F^0_\alpha$, bottom)
with electron flavor (left) and $x$ flavor (right). In the latter case,
we plot the sum of muonic and tauonic fluxes. 
All fluxes refer to our reference supernova model and to a distance $d=10$~kpc. The
black, green, blue and magenta curves correspond to 
$t = 1$, 5, 10, and 20~s, respectively.
\label{fig1}}
\end{figure}
%%%%%%%%%%%%%%%%%%%%%%%%%%%%%%%%%%%%%%%%%%%%%%%%%%%%%%%%%%%%%%%%%%%%%%

\subsection{Supernova model}

Our reference SN scenario is essentially taken from \cite{Schir}, with minor changes in the average neutrino energies.
We assume a galactic core-collapse supernova releasing a binding energy $E_B=3\times 10^{53}$~erg,
equally distributed among the six ($3\nu+3\overline\nu$) species, and distributed in time
with a time constant $\tau=3$~s. The  energy luminosity associated to each species is thus
%............................
\begin{equation}
\label{lum}
L(t)=\frac{E_B}{6}\frac{e^{-t/\tau}}{\tau}\ .
\end{equation}
%............................
At the conventional distance for a galactic supernova,
\begin{equation}
d=10\mathrm{\ kpc}\ ,
\end{equation} 
the unoscillated flux of the neutrino species $\nu_\alpha$, 
per unit of area, time, and energy, is expected to be
%...................................................................
\begin{equation}
\label{flux}
F^0_\alpha(E,\,t) = \frac{L(t)}{4\pi d^2}\,\frac{\phi^0_\alpha(E)}{\langle E_\alpha\rangle}\ ,
\end{equation}
%...................................................................
where we assume normalized thermal energy spectra $\phi^0_\alpha(E)$ 
with average energy $\langle E_\alpha \rangle$. As in \cite{Miri}, inspired by \cite{Totani}, we set
%..................
\begin{equation}
\langle E_e\rangle=10~\mathrm{MeV},\  
\langle \overline E_{ e}\rangle=15~\mathrm{MeV},\  
\langle E_{x} \rangle=24~\mathrm{MeV}\ ,
\label{refenergies}
\end{equation}
%.....................
corresponding to temperatures  $(T_e,\,T_{\bar e},\,T_x)=
(3.17,\, 4.76,\, 7.62)$ MeV.%
%------
\footnote{Effects of variations with respect to the default choice in Eq.~(\protect\ref{refenergies})
will be discussed in Sec.~6.}
%---------
 Figure~1 shows the corresponding fluxes $F^0_\alpha(E,\,t)$ in the
energy interval $E\in [1,50]$~MeV, at four representative times ($t=1$, 5, 10, and 20~s).

Concerning the SN geometry, we adopt the spherically-symmetric bulb model \cite{Semi},
with a neutrinosphere radius $R_\nu=10$~km. Within this model, the previous initial conditions
for the neutrino luminosity and unoscillated fluxes define,  at any radius $r>R_\nu$, 
the effective density of neutrinos 
 ($N=N_e+N_\mu+N_\tau$) and of antineutrinos 
($\overline N$) per unit volume, as well as the self-interaction   
potential
%........................................................
\begin{equation}
\mu(r) = \sqrt{2}\, G_F\, [N(r)+\overline N(r)]\ . \label{mu}
\end{equation}
%.........................................................
For each species $\nu_\alpha$, it is useful to introduce also 
density $n_\alpha$ per unit of volume and energy (see \cite{Semi,Miri} for details):
%........................................................
\begin{equation}
N_\alpha = \int dE\,n_\alpha(E)\ .
\end{equation}
%.........................................................

Figure~2 (left panel) shows the function $\mu(r)$, up to $r=500$~km, 
for the four $t$ values chosen. In the same panel, 
the shaded horizontal band corresponds to the range
$\mu\in [\mu_\mathrm{inf},\,\mu_\mathrm{sup}]\simeq [6.7,\,67]$~km$^{-1}$ 
where, according to the estimate in \cite{Miri},
collective bipolar \cite{Bipol} (``pendulum'' \cite{Pend}) oscillations are expected to develop 
(after a ``synchronized'' regime \cite{Bipol,Pend,Sync}) for this SN
scenario in the $2\nu$ case. 
We have explicitly verified that these expectations 
also hold in our full $3\nu$ flavor evolution, with a radial accuracy better than $\sim\!10$~km (not shown).
%..................................
For example, at $t=10$~s, we find numerically a bipolar development range $r\simeq [60,\,100]$~km, in
good agreement with the range $r\simeq [55,\,95]$~km derived
from the intersection of the horizontal band with the $\mu(r)$ curve at $t=10$~s.

%%%%%%%%%%%%%%%%%%%%%%%%%%% FIGURE 2 %%%%%%%%%%%%%%%%%%%%%%%%%%%%%%%%
\begin{figure}[t]
\centering
\vspace*{4mm}
\hspace*{22mm}
\epsfig{figure=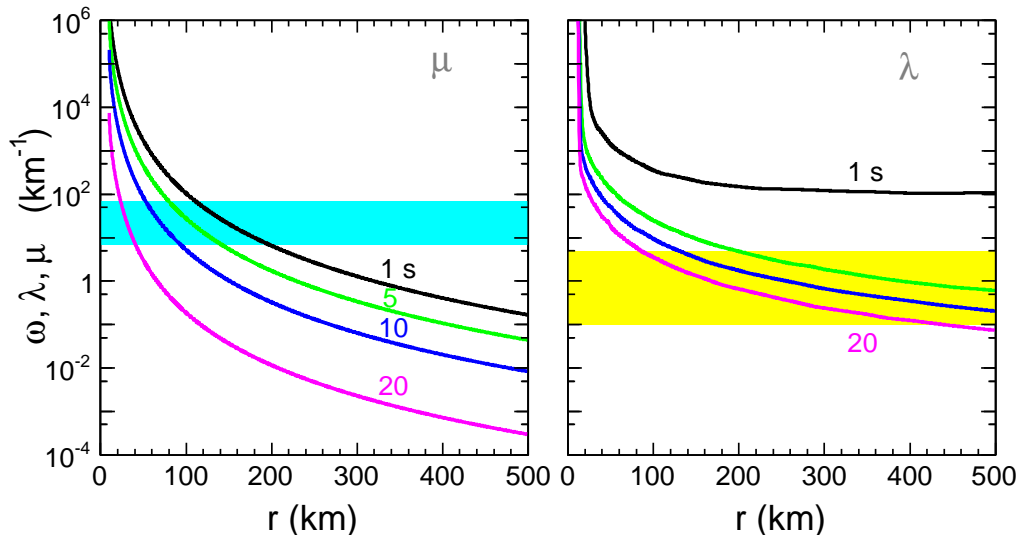,width=0.85\columnwidth}
\vspace*{-6mm}
\caption{Radial profiles of the
self-interaction potential $\mu$ (left) and of the matter potential $\lambda$ (right). The
black, green, blue and magenta curves correspond to 
$t = 1$, 5, 10, and 20~s, respectively. In the left panel, the shaded horizontal band
marks the $\mu$ range where bipolar effects develop. In the right panel, the 
shaded band marks the range of the high vacuum frequency $\omega_H$ for $E\in[1,\,50]$~MeV, where
MSW effects may develop if the $H$-resonance condition ($\lambda \sim \omega_H$) is satisfied. 
\label{fig2}} 
\end{figure}
%%%%%%%%%%%%%%%%%%%%%%%%%%%%%%%%%%%%%%%%%%%%%%%%%%%%%%%%%%%%%%%%%%%%%%

However, collective phenomena extend somewhat beyond the bipolar range,
and eventually vanish when the spectral splits are complete. As a rule of thumb, 
the end of all collective effects occurs at a radius $r_\mathrm{end}$ where 
the self-interaction potential $\mu$ equals a relatively small
value $\mu_\mathrm{end}$.
In our numerical experiments, we find that $\mu_\mathrm{end}\simeq 0.5$~km$^{-1}$ 
provides a reasonable estimate for the radius $r_\mathrm{end}$
%.........
\begin{equation}
\label{rend}
\mu(r_\mathrm{end})\simeq 0.5$~km$^{-1}\ , 
\end{equation}
%...........
at any time $t$ (e.g., $r_\mathrm{end}\simeq 400$~km at $t=1$~s in Fig.~2). 
For safety, we have always checked that our results do not appreciably change 
by continuing the numerical flavor evolution for further $\sim\! 100$ km or more 
(e.g., up to $r\simeq 500$~km for $t=1$~s).

Figure~2 (right panel) shows the radial profile of the matter potential $\lambda$ at different times, 
as taken from \cite{Schir}. As usual, the matter potential is defined as 
%........................................................
\begin{equation}
\lambda(r) = \sqrt{2}\, G_F\, N_{e^-}(r)\ , 
\label{lambda}
\end{equation}
%.........................................................
where $N_{e^-}(r)$ is the net electron density. In inverted hierarchy, 
antineutrinos may undergo significant matter effects
(or MSW effects, from Mikheev, Smirnov, and Wolfenstein \cite{Matt}) when 
the so-called $H$-resonance condition is fulfilled (see, e.g., \cite{Hres}),
%........................................................
\begin{equation}
\lambda \simeq \omega_H\ .
\label{res_H}
\end{equation}
%.........................................................
For the energy range in Fig.~1 ($E\in[1,\,50]$~MeV), the relevant interval 
for the high vacuum frequency, $\omega_H\in [0.1,\,5]$~km$^{-1}$, is shown 
in the right panel of Fig.~2 as a shaded horizontal band.  
[Note in Fig.~2 that the $\omega_H$ value of
0.5~km$^{-1}$ in Eq.~(\ref{rend}) is in the middle of such band.]
Its intersection 
 with one of the $\lambda(r)$ curves marks the related $H$-resonance range 
 (for example, from Fig.~2, $r\gtrsim 130$~km at $t=10$~s).

A comparison of the left and right panels in Fig.~2 shows that, in our reference SN model, 
collective bipolar oscillations develop before possible $H$-resonance effects
at any time (while they may be coupled in other scenarios with shallower matter profiles, see e.g.\
 \cite{Semi,Basu}). Flavor swap effects, however, extend up to $r_\mathrm{end}$ estimated previously, and
may partly enter the $H$-resonance region for large $\omega_H$ at $t>1$. However, for values
of $\theta_{13}$ (at least) as small as in  Eq.~(\ref{theta13}), the $H$ resonance is 
highly nonadiabatic, and produces no further flavor conversion at any relevant SN energy \cite{Hres}. 
We have numerically verified the absence of MSW flavor conversion for antineutrinos 
in inverted hierarchy (besides their collective flavor swap) for $r$ well within the $H$-resonance region.
We have also verified that all our main results in Secs.~4 and~5 remain unchanged
for reasonably smaller values of $\sin^2\theta_{13}$ (e.g., $10^{-7}$ or $10^{-8}$), which
simply cause a logarithmic delay of the bipolar oscillation onset \cite{Coll,Pend}. 
As an added bonus, for $\sin^2\theta_{13}\lesssim 10^{-6}$, one gets a basically
complete suppression of: ($i$) possible matter effects due to
shock-wave features (see, e.g., \cite{Shock}); ($ii$)  
possible decoherence effects due to density fluctuations (see, e.g., \cite{Turb}); 
($iii$) Earth matter effects, if any, before detection (see, e.g., \cite{Di08,Earth}).
These welcome simplifications make neutrino-neutrino interaction effects 
dominate over neutrino-matter interaction effects
at any time $t\geq 1$~s in our framework,
and motivate {\em a posteriori\/} our choice for the tiny mixing angle in Eq.~(\ref{theta13}).

A final remark is in order. As described in the next Sections, we evolve the neutrino flavors along
a single, averaged radial trajectory. This (so-called ``single-angle'') approximation is often reasonable---when 
compared with few available ``multi-angle'' calculations---but it may fail badly when the matter 
potential is so high that its gradient depends sensitively on the trajectory  (see
\cite{Multi}, and references therein). For our inspiring SN model 
\cite{Schir},
this situation may occur only at very early stages ($t<1$~s), which are excluded from the present 
investigation.

%%%%%%%%%%%%%%%%%%%%%%%%%%%%%%%%%%%%%%%%%%%%%%%%%%%
\section{Three-flavor formalism}

The density matrix formalism is particularly useful to deal with neutrino self-interactions, which
depend on the neutrino density itself. In the two-family case, the traceless part
of the ($2\times 2$) density matrix \mbox{\boldmath$\rho$} can be expanded onto Pauli matrices $\sigma_i$, with projections
coefficients forming a 3-vector $\mathbf{P}$ for each energy mode $E$ \cite{Pend}.

In the three-family case, we have generalized the $2\nu$ evolution equations to the $3\nu$ case
\cite{Laurea},
by projecting  the ($3\times 3$)  density matrix \mbox{\boldmath$\rho$} in flavor 
basis onto Gell-Mann matrices $\lambda_i$, in terms of an 8-component vector $\mathbf{P}$.
The resulting 
equations  \cite{Laurea} are formally similar to those recently discussed in 
\cite{Das3,Full3,Basu}, and are briefly reviewed below for completeness.

Given an 8-dimensional orthonormal basis $(\mathbf{e}_1,\dots,\mathbf{e}_8)$ for the $\mathbf{P}$ vector space,
the flavor projector matrices, $\mbox{\boldmath$\Pi$}_e=\mathrm{diag}(1,\,0,\,0)$, 
 $\mbox{\boldmath$\Pi$}_\mu=\mathrm{diag}(0,\,1,\,0)$ and  $\mbox{\boldmath$\Pi$}_\tau=\mathrm{diag}(0,\,0,\,1)$,
can be represented as
%..........................
\begin{equation}
\mbox{\boldmath$\Pi$}_\alpha = \frac{\mathbf{1}}{3}+
\mathbf{u}_\alpha \cdot \frac{\mbox{\boldmath$\lambda$}}{2} \ \ (\alpha=e,\,\mu,\,\tau)\ ,
\end{equation}
%...........................
where $\mathbf{1}$ is the $3\times 3$ unit matrix, $\mbox{\boldmath$\lambda$}=\sum \lambda_i\, \mathbf{e}_{i}$, and
the $\lambda_i$'s are the Gell-Mann matrices with usual conventions: $\lambda_i=\lambda_i^\dagger$; tr$(\lambda_i)=0$;
tr$(\lambda_i\,\lambda_j)=2\delta_{ij}$; and $[\lambda_i,\,\lambda_j]=2if_{ijk}\lambda_k$. The $\mathbf{u}_\alpha$
vectors (with $|\mathbf{u}_\alpha|=2/\sqrt{3}$) read:
%..........................
\begin{equation}
\mathbf{u}_e= \mathbf{e}_3+\frac{1}{\sqrt{3}}\mathbf{e}_8\ ,\ 
\mathbf{u}_\mu= -\mathbf{e}_3+\frac{1}{\sqrt{3}}\mathbf{e}_8\ ,\ 
\mathbf{u}_\tau= -\frac{2}{\sqrt{3}}\mathbf{e}_8\ . 
\end{equation}
%...........................

The neutrino density matrix, at given energy $E$, is projected via 
%.........................
\begin{equation}
\label{matitre}
 \mbox{\boldmath$\rho$} =  n\left(\frac{\mathbf{1}}{3}+
\mathbf{P} \cdot \frac{\mbox{\boldmath$\lambda$}}{2}\right)\ ,
\end{equation}
%.........................
where $\mathbf{P}=\sum P_i\,\mathbf{e}_i$ (with $|\mathbf{P}|=2/\sqrt{3}$), and 
$n$ is the neutrino density per unit volume and energy introduced before. At any time $t$ and
energy $E$,
the $\nu_\alpha$ content  is given by tr$({\mbox{\boldmath$\rho$}}(t)
{\mbox{\boldmath$\Pi$}}_\alpha$).

The evolution of \mbox{\boldmath$\rho$} is governed by the Liouville equation 
$i \,\dot{\mbox{\boldmath$\rho$}}=[\mathbf{H},\,{\mbox{\boldmath$\rho$}}]$, where
the Hamiltonian $\mathbf{H}$ contains vacuum, matter, and self-interaction terms. 
In the vacuum term, we include both  low and high vacuum frequencies ($\omega_L$ and
$\omega_H$, respectively), and assume inverted hierarchy. In the matter term, we include 
the $\nu_\tau-\nu_\mu$ potential difference at one loop \cite{Botella},  whose size is
$\delta\lambda/\lambda\simeq 5\times 10^{-5}$  (see also \cite{Esteban,Full3}). By expanding 
\mbox{\boldmath$\rho$} and $\mathbf{H}$ onto Gell-Mann matrices, evolution equations are obtained
for the $\nu$ flavor polarization vector $\mathbf{P}$ and, analogously, for the
$\overline\nu$ vector $\overline\mathbf{P}$:
%.........................
\begin{eqnarray}
\label{evol3nu} \dot{\mathbf{P}} &=& [+ (\omega_{L} \mathbf{B}_L - \omega_{H} \mathbf{B}_H)+
 \lambda\mathbf{v}  + \mu \mathbf{D}] \times
\mathbf{P}\ ,\\
\label{evol3antinu} \dot{\overline{\mathbf{P}}} &=& [- (\omega_{L} \mathbf{B}_L - \omega_{H} \mathbf{B}_H)+
 \lambda\mathbf{v}  + \mu \mathbf{D}] \times
\overline{\mathbf{P}}\ ,
\end{eqnarray}
%...........................
where the 8-component vector product is defined as $(\mathbf{a}\times\mathbf{b})_i=f_{ijk}\,a_j\,b_k$.

In the above equations, the first (vacuum) terms embed the squared mass splittings via $\omega_{L,H}$, and
the mixing angles via effective ``magnetic fields'' $\mathbf{B}_{L,H}$ with components
%.........................
\begin{equation}
\label{BL} 
\mathbf{B}_L = \left(
\begin{array}{c}
c_{13} (  S_{12}c_{23} + C_{12} s_{13} s_{23})\\ 
0\\
-\frac{1}{8}C_{12} ( 3 + 3C_{13} + 3C_{23} -C_{13}C_{23})  + \frac{1}{2} s_{13} S_{12}  S_{23}\\
\frac{1}{2}  C_{12}  S_{13} c_{23} - S_{12} c_{13} s_{23}  \\ 
0\\
- s_{13}C_{23} S_{12}   -  \frac{1}{4} (3 - C_{13}) S_{23} C_{12}\\  
0\\ 
-\frac{\sqrt{3}}{8} C_{12}(1+C_{13}-3C_{23}+C_{13}C_{23}) -\frac{\sqrt{3}}{2}s_{13}S_{12}S_{23}
\end{array}
\right)\ ,
\end{equation}
%.........................
and
%.........................
\begin{equation}
\label{BH} 
\mathbf{B}_H = \left(
\begin{array}{c}
S_{13}s_{23}\\
0\\
\frac{1}{4}(1-3C_{13}+C_{23}+C_{13}C_{23})\\
S_{13}c_{23}\\
0\\
\frac{1}{2}S_{23}(1+C_{13})\\
0\\
\frac{1}{4\sqrt{3}}(1-3C_{13}-3C_{23}-3C_{13}C_{23})
\end{array}
\right)\ ,
\end{equation}
%.........................
where we have set $s_{ij}=\sin\theta_{ij}$, $c_{ij}=\cos\theta_{ij}$, $S_{ij}=\sin2\theta_{ij}$, and
$C_{ij}=\cos2\theta_{ij}$.

The second (matter interaction) term $\lambda \mathbf{v}$ in
Eqs.~(\ref{evol3nu}) and (\ref{evol3antinu}) embeds the $\nu_e- \nu_{\mu,\tau}$ interaction
energy difference (oriented along $\mathbf{u}_e$), as well as the tiny correction due
the $\nu_\tau-\nu_{\mu}$ interaction energy difference (oriented along $\mathbf{u}_\tau$), 
%..........................
\begin{equation}
\lambda \mathbf{v}=\lambda\left(\mathbf{u}_e+\frac{\delta \lambda}{\lambda} \mathbf{u}_\tau\right)\ .
\end{equation}
%..........................
The third (self-interaction) term $\mu\mathbf{D}$ couples all neutrino and antineutrino modes via the integral 
vector difference
%..........................
\begin{equation}
\mathbf{D} 
= \frac{1}{N + \overline{N}} \int dE\  (n\,\mathbf{P} - \overline n\,\overline{\mathbf{P}})\ .
\end{equation}
%.....................
Note that, in general, neutrino self-interactions depend on the crossing angle between the 
neutrino trajectories \cite{Pant,Semi}. We have implicitly assumed that such dependence can be averaged
out along a single, radial trajectory 
(single-angle approximation, see \cite{Miri} and references therein). 
If crossing angles 
were accounted for, the equations would entail a further angular variable 
(multi-angle description) \cite{Semi,Nonl,Miri,Estmulti} not included in the present investigation.

We discretize in energy the coupled evolution equations (\ref{evol3nu}) and 
(\ref{evol3antinu}),  and solve them
by numerical integration (see \cite{Miri} for details), up to the end of collective
effects ($r_\mathrm{end}\lesssim 500$~km). The 
initial conditions are fixed by the SN model described in 
the previous Section, for each representative time: $t=1$, 5, 10 and 20~s. 
The results are described below in the graphically convenient range $E\in[1,\,50]$~MeV
(although the numerical evolution includes modes with $E<1$~MeV).

\section{Intermediate fluxes after collective effects}

In this Section we discuss our numerical results for the intermediate 
fluxes $F'_\alpha$ at the end
of collective effects ($r=r_\mathrm{end}\lesssim 500$~km), and show that they confirm 
the expectations in Eqs.~(\ref{je})--(\ref{jxbar}). 
In order to allow a visual comparison
with the unoscillated fluxes $F^0_\alpha$ at $d=10$~kpc (Fig.~1), the fluxes $F'_\alpha$ are rescaled by 
a factor $(r_\mathrm{end}/d)^2$. 
We stress that the intermediate fluxes $F'_\alpha$ are unobservable, as both $\nu$ and $\overline\nu$ are subject
to further flavor evolution in the SN, and to phase-averaging effects up to the detector. The final, observable
fluxes $F_\alpha$ at $d=10$~kpc will be estimated in the next Section.

%%%%%%%%%%%%%%%%%%%%%%%%%%% FIGURE 3 %%%%%%%%%%%%%%%%%%%%%%%%%%%%%%%%
\begin{figure}[t]
\centering
\vspace*{4mm}
\hspace*{22mm}
\epsfig{figure=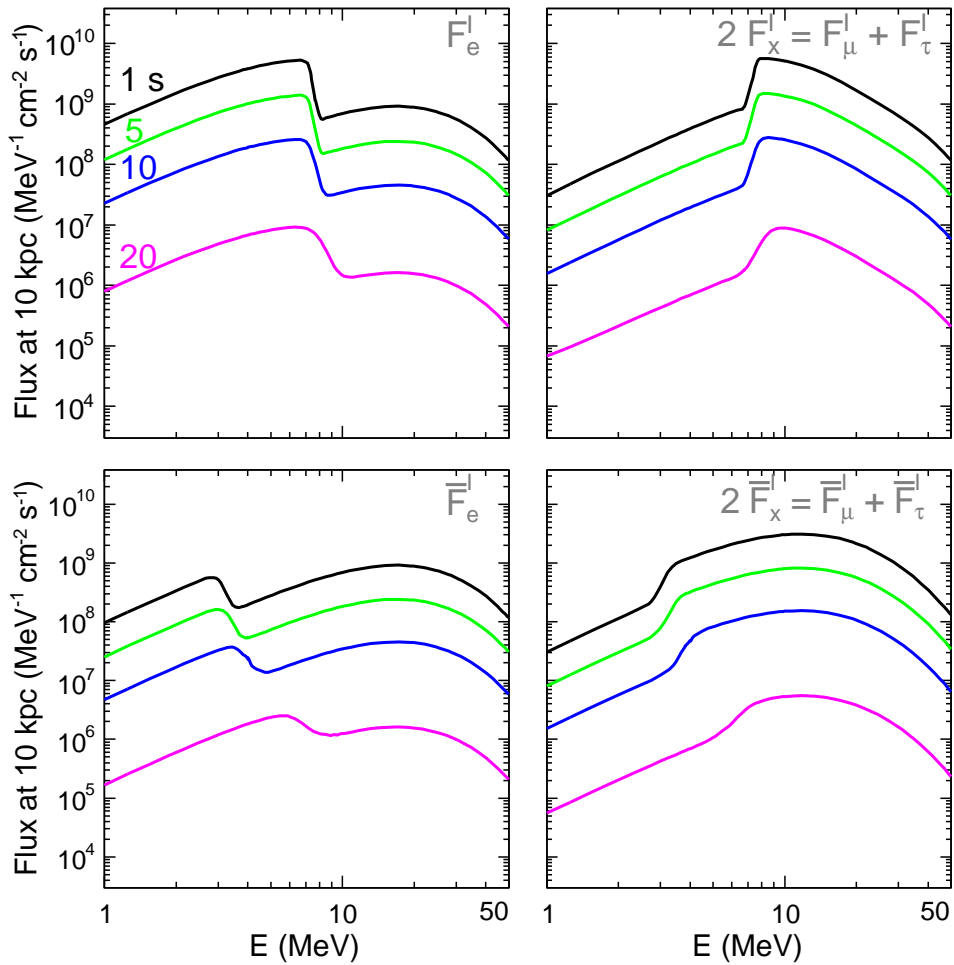,width=0.8\columnwidth}
\vspace*{0mm}
\caption{Fluxes of neutrinos ($F'_\alpha$) and antineutrinos
($\overline F'_\alpha$) at the end of collective effects, 
rescaled to $d=10$~kpc for comparison with Fig.~1. See the text for details. 
\label{fig 3}}
\end{figure}
%%%%%%%%%%%%%%%%%%%%%%%%%%%%%%%%%%%%%%%%%%%%%%%%%%%%%%%%%%%%%%%%%%%%%%

Figure~3 shows our computed fluxes $F'_\alpha$, in the same
scale as Fig.~1. The comparison of Figs.~1 and 3 confirms that spectral splits of neutrinos and antineutrinos
emerge as dominant features, as expected. In our adopted scenario,
the neutrino split occurs around a critical energy $E_c\simeq 7$~MeV \cite{Miri}, 
dictated by lepton number conservation \cite{Simple,Smirnov}. The antineutrino split occurs at a somewhat lower energy
$\overline E_c\simeq \mathrm{few}$~MeV, which is not predicted a priori so far \cite{Tamb}. 
However, apart from a transition region (about $1$--2~MeV wide)
around $E_c$ and $\overline E_c$, the evolved fluxes in Fig.~3 coincide 
with those expected from the application of Eqs.~(\ref{je})--(\ref{jxbar})
to the unevolved fluxes of Fig.~1.

 %%%%%%%%%%%%%%%%%%%%%%%%%%% FIGURE 4 %%%%%%%%%%%%%%%%%%%%%%%%%%%%%%%%
\begin{figure}[t]
\centering
\vspace*{4mm}
\hspace*{22mm}
\epsfig{figure=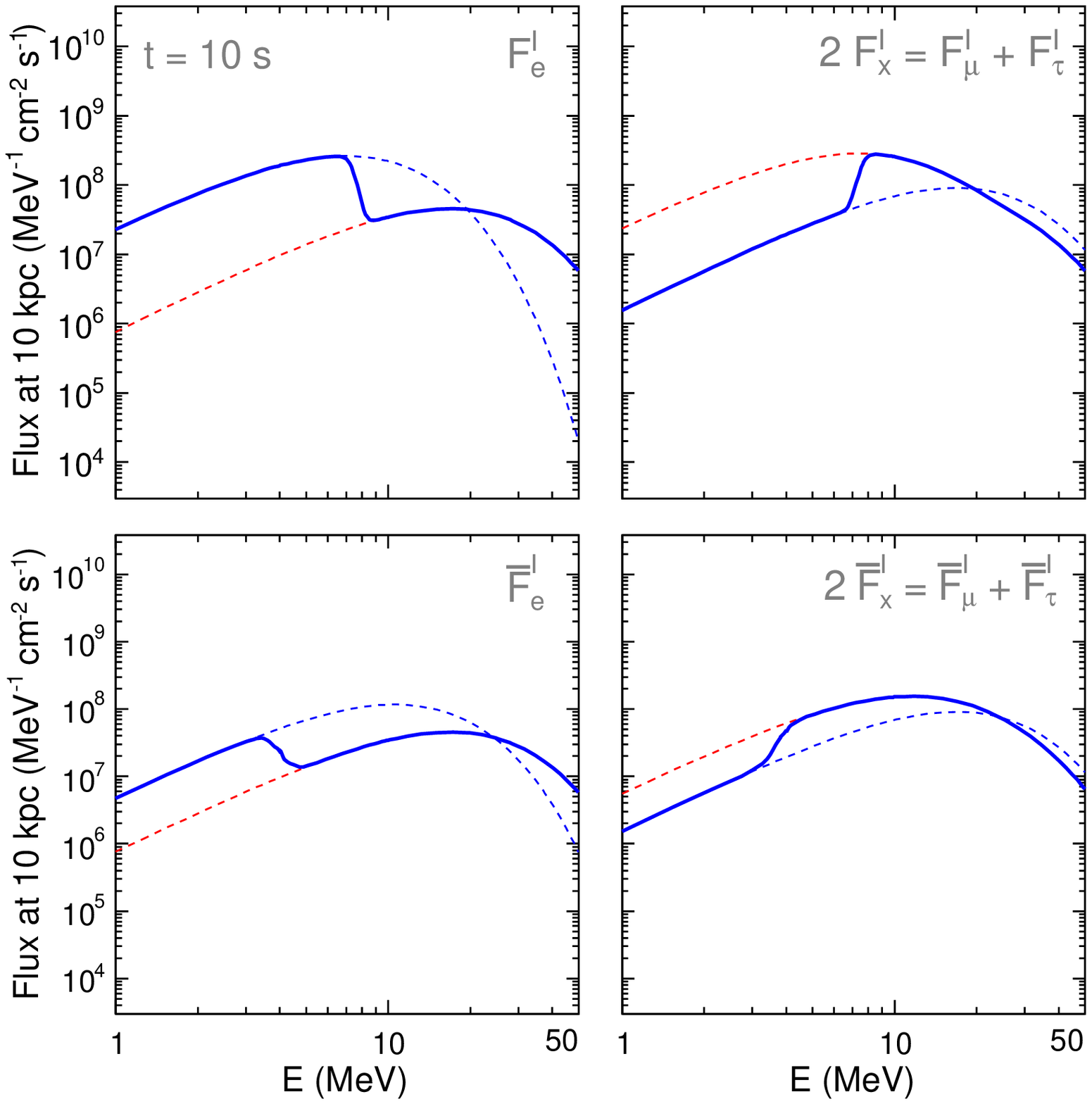,width=0.8\columnwidth}
\vspace*{0mm}
\caption{Fluxes of neutrinos and antineutrinos at the end of collective effects, for
$t = 10$~s. Solid blue curves: computed fluxes $F'_\alpha$. 
Dashed blue and red curves: limiting behavior at low-energy and high-energy, respectively, in terms of 
unoscillated fluxes $F^0_\alpha$, according to Eqs.~(\ref{je})--(\ref{jxbar}). 
\label{fig 4}}
\end{figure}
%%%%%%%%%%%%%%%%%%%%%%%%%%%%%%%%%%%%%%%%%%%%%%%%%%%%%%%%%%%%%%%%%%%%%%

This is shown in more detail in Fig.~4, with reference to the time $t=10$~s (similar results hold  at
1, 5, and 20~s). Below the critical energy, all  $\nu_\alpha$ species remain unchanged:  
the evolved fluxes $F'_\alpha$ (solid blue curves) coincide with the unevolved
fluxes $F^0_\alpha$ (dashed blue curves) in the upper left panel. 
Conversely, above the critical energy, the electron flavor fully converts. As a consequence,
in the lower left panels of Fig.~4, the
evolved fluxes $F'_e$ and $\overline F'_{e}$  turn into $F^0_x$ and $\overline F^0_{x}$, respectively 
(dashed red curves). Antineutrinos show a similar behavior (right panels).
Therefore, Eqs.~(\ref{je})--(\ref{jxbar}) hold with good accuracy, except 
close to the spectrum step, whose details may also depend on the adopted single-angle approximation.
Indeed, multi-angle calculations (not performed in this work) may further widen the transition region 
\cite{Miri}, especially for antineutrinos \cite{Tamb}. However, they are not expected to
change the low- and high-energy limits of the spectra. 

%%%%%%%%%%%%%%%%%%%%%%%%%%% FIGURE 5 %%%%%%%%%%%%%%%%%%%%%%%%%%%%%%%%
\begin{figure}[t]
\centering
\vspace*{4mm}
\hspace*{22mm}
\epsfig{figure=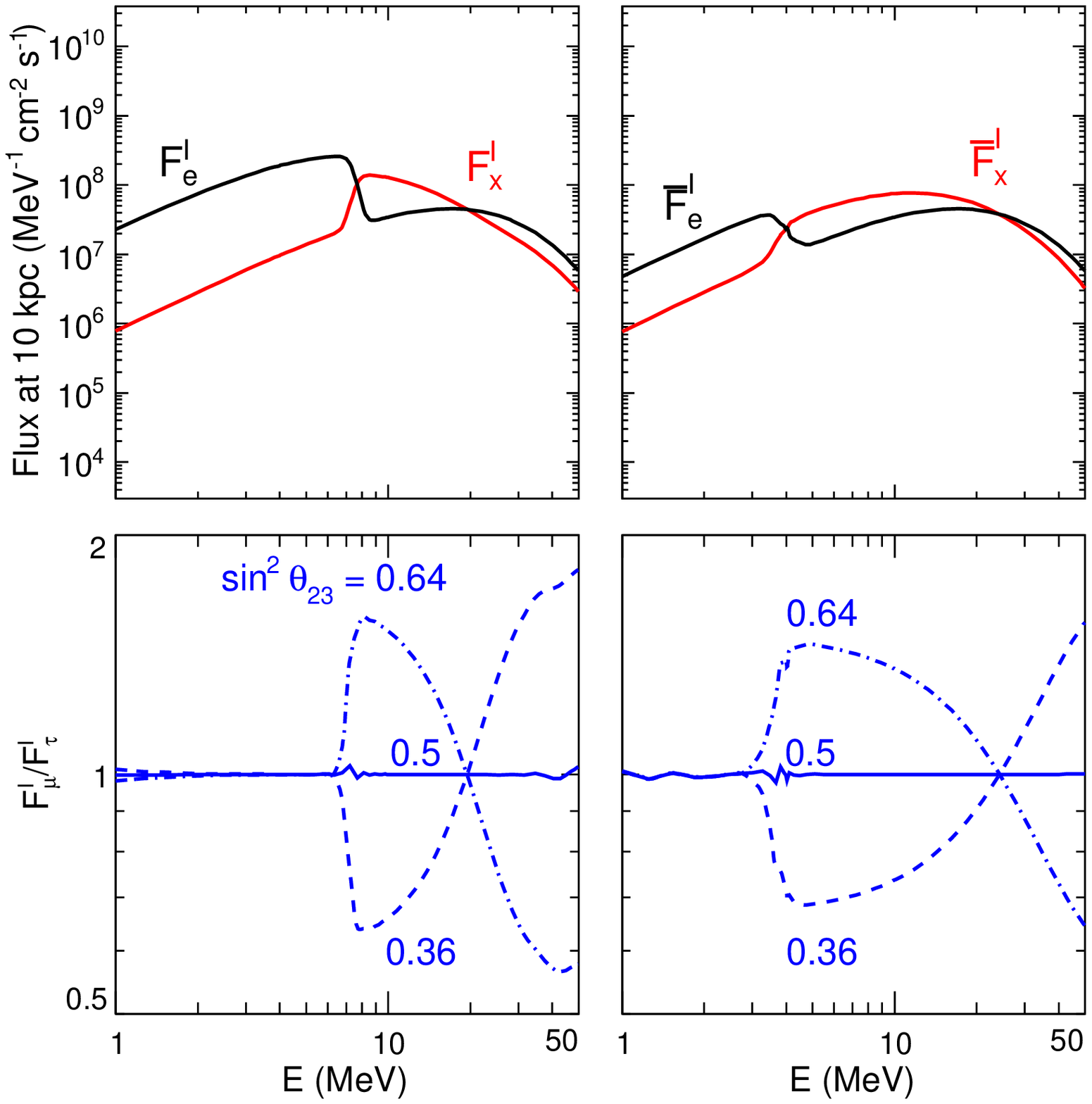,width=0.8\columnwidth}
\vspace*{0mm}
\caption{Neutrino and antineutrino fluxes $F'_\alpha$ at the end of collective effects ($r\lesssim 500$~km), 
for $t=10$~s. Upper panels: absolute fluxes. Lower panels: muonic-to-tauonic flux ratio for $\nu$
(left) and $\overline\nu$ (right), for $\sin^2 \theta_{23} = 0.36,\, 0.5,\, 0.64$.
\label{fig 5}}
\end{figure}
%%%%%%%%%%%%%%%%%%%%%%%%%%%%%%%%%%%%%%%%%%%%%%%%%%%%%%%%%%%%%%%%%%%%%%

The above  $3\nu$ results, obtained at the end of collective effects,
are not significantly influenced by 
the subdominant ``solar'' squared mass difference ($\delta m^2$) or by the $\nu_{\mu}-\nu_\tau$ interaction energy
difference ($\delta \lambda$). Effects of $\delta m^2$  
are expected to be larger at relatively shallow matter densities
\cite{Full3,Basu},
i.e., at larger $t$ in Fig.~2; we find the largest fractional flux variations at $t=20$~s, as obtained
by setting $\delta m^2=0$, to be negliglible ($<\mathrm{few}\%$). Conversely,
effects of $\delta\lambda$ are expected to be larger at higher matter densities \cite{Esteban,Full3,Gava}, 
i.e., at shorter $t$; we
find the largest variations at $t=1$~s, as obtained by setting $\delta\lambda/\lambda=0$, to be also negligible
($<\mathrm{few}\%$). We find that 
these variations are mainly localized around the critical split energies, 
 and can thus be mainly attributed to small, subleading $3\nu$ collective effects.

To a very good accuracy, our results are independent of any effect which
may change the relative $\nu_\mu$ and $\nu_\tau$ flux proportions within the sum
$2F'_x=F'_e+F'_\mu$, including not only the
small $\nu_{\mu,\tau}$ correction $\delta\lambda/\lambda$, but also the (quasi)maximal
$\nu_{\mu,\tau}$ mixing angle $\theta_{23}$. Indeed, the curves in Fig.~3 
do not depend on the chosen value of $\sin^2\theta_{23}$ in Eq.~(\ref{theta23}), as we have explicitly 
checked.

For instance, Fig.~5 shows both the absolute fluxes with
$e$ and $x$ flavor (upper panels) and the muonic-to-tauonic flavor ratio (lower panels) at the
end of collective effects, for $t=10$~s; similar results hold at different $t$ (not shown). 
The curves in the upper panels do not depend on the indicated value of
$\sin^2\theta_{23}$, while those in the lower one do. More precisely,
for neutrinos (lower left panel) the ratio $F'_\mu/F'_\tau$ remains equal to $1(=F_\mu^0/F_\tau^0)$ 
in three cases: ($i$) at maximal mixing, $\theta_{23}=\pi/4$, where
$\nu_\mu$ and $\nu_\tau$ are interchangeable, up to minor $\delta\lambda/\lambda$ effects; ($ii$)
below the critical energy $E_c\simeq 7$~MeV, where there is no net flavor conversion; and ($iii$)
at the specific ``equalization'' energy $E_\mathrm{eq}\simeq 20$~MeV where $F^0_e=F^0_x$,
and flavor conversions become inoperative: $F'_e=F'_x$. 
Similarly for antineutrinos (lower right panel), but with $\overline E_c\simeq 3$--4~MeV
and $\overline E_\mathrm{eq}\simeq 27$~MeV. 

In Fig.~5, the cases where $F'_\mu/F'_\tau\neq 1$ can be understood
by considering that, in inverted hierarchy, flavor conversions  occur mainly between the 
higher mass state $\nu_3$ and the lower mass doublet $\nu_{1,2}$, with $\nu_3$ being richer 
in $\nu_\tau$ or $\nu_\mu$ according to $\theta_{23}$ being in the first or second octant
(while the $\nu_{1,2}$ doublet is always richer in $\nu_e$). If
$F^0_e>F^0_{\mu,\tau}$, namely $E<E_\mathrm{eq}$, then $\Delta m^2$-driven conversions are dominantly
of the kind $\nu_e\to \nu_\tau$ ($\nu_e\to\nu_\mu$)
for $\sin^2\theta_{23}<1/2$ ($\sin^2\theta_{23}>1/2$),
so that the final $\nu_\tau$ ($\nu_\mu$) flavor increases at the expenses of $\nu_\mu$ ($\nu_\tau$).
The opposite happens if $F^0_e<F^0_{\mu,\tau}$, namely $E>E_\mathrm{eq}$. The same reasoning hold
for antineutrinos.

These arguments explain the main qualitative features of the numerical
results in the lower panels of Fig.~5 which, by themselves, have just an academic interest. 
They may serve, however, as benchmarks in the exploration of extended SN scenarios with 
nonstandard initial conditions \cite{Luna} or interactions \cite{Blennow} in the $\nu_\mu$-$\nu_\tau$ sector.

%%%%%%%%%%%%%%%%%%%%%%%%%%%%%%%%%%%%%%%%%%%%%%%%%%%%%%%%%%%
\section{Final fluxes at the detector}

After collective effects have ended, and the $H$-resonance region is traversed,  the flavor evolution
is eventually subject to the so-called ``$L$ resonance'' at $\lambda\sim \omega_L$,
which can be assumed  adiabatic (see, e.g., \cite{Hres}). 
[As noted, additional effects due
to possible shock wave features or density fluctuations are 
ineffective for $\theta_{13}$ as low as in Eq.~(\ref{theta13}).] 

In the specific
context of self-interacting SN neutrinos in three families, 
final $L$-resonance effects have also been numerically verified in \cite{Das3},
by continuing the numerical evolution up to a few thousand km. Here we do not repeat this numerical
check of long-distance evolution, but take the resulting effects for granted; for completeness,
we rephrase the related arguments of \cite{Das3} in our notation as follows. 

In the absence of self interactions effects,
a strongly nonadiabatic $H$ resonance, plus an adiabatic $L$ resonance, 
would eventually distribute the original fluxes over the effective mass eigenstates $\nu_i$ and $\overline\nu_i$ as
\cite{Hres}: 
%............................
\begin{eqnarray}
 F_1&=&F_x^0\ ,\\
 F_2&=&F_e^0\ , \\
 F_3&=&F_x^0\ , \\
  \overline F_1&=&\overline F_e^0\ ,\\ 
  \overline F_2&=&\overline F_x^0\ , \\ 
  \overline F_3&=&\overline F_x^0\ .
\end{eqnarray}
%...............................
Spectral splits, however, alter these standard MSW expectations and 
swap the flavor contents in the effective $(e,\,3)$ sector \cite{Das3}, so that
%............................
\begin{eqnarray}
\label{first}
 F_1&=&F_x^0\ ,\\
 F_2&=&F_e^0P_{ee}+F_x^0(1-P_{ee})\ , \\
 F_3&=&F_x^0P_{ee}+F^0_e(1-P_{ee})\ , \\
  \overline F_1&=&\overline F_e^0 \overline P_{ee}+\overline F_x^0 (1-\overline P_{ee})\ ,\\ 
  \overline F_2&=&\overline F_x^0\ , \\ 
  \overline F_3&=&\overline F_x^0 \overline P_{ee}+\overline F_e^0 (1-\overline P_{ee})\ ,
\label{last}
\end{eqnarray}
%...............................
where $P_{ee}$ and $\overline P_{ee}$, embedding collective effects, have a stepwise behavior 
as in Eqs.~(\ref{P}) and (\ref{Pbar}). 
The final, phase-averaged fluxes for the electron flavor at detection are given by
%..............................
\begin{eqnarray}
\label{lin1}
F_e &=& \sum |U_{ei}|^2 F_i \simeq \cos^2\theta_{12}F_1 + \sin^2\theta_{12} F_2\ ,\\
\label{lin2}
\overline F_e &=& \sum |U_{ei}|^2 \overline F_i \simeq \cos^2\theta_{12}\overline F_1 + \sin^2\theta_{12} \overline F_2\ ,
\end{eqnarray}
%...............................
which, together with Eqs.~(\ref{first})--(\ref{last}), 
reproduce the limiting behaviors discussed in \cite{Das3},
%...............................
\begin{eqnarray}
F_e &\simeq & \left\{\begin{array}{ll}\cos^2\theta_{12} F^0_x + \sin^2\theta_{12}F^0_{e} 
& (E<E_c)\ ,\\ F^0_x & (E>E_c)\ ,\end{array} \right.\\
\overline F_e &\simeq & \left\{\begin{array}{ll}\sin^2\theta_{12}\overline F^0_{x} + \cos^2\theta_{12}\overline F^0_e  
& (E<\overline E_c)\ ,\\ \overline F^0_x & (E>\overline E_c)\ ,\end{array} \right.
\end{eqnarray}
%..............................
where the low-energy $\overline \nu$ split, not considered in \cite{Das3}, is also included.

By using Eqs.~(\ref{je})--(\ref{jxbar}), one can eliminate $P_{ee}$ and $\overline P_{ee}$
in terms of the fluxes $F'_\alpha$ at the end of collective effects. 
The final fluxes $F_e$ and $\overline F_e$ can thus be expressed  
in terms of the initial fluxes in Fig.~1 and of the intermediate computed fluxes
in Fig.~3,
%...............................
\begin{eqnarray}
F_e &=& \cos^2\theta_{12} F_x^0+\sin^2\theta_{12}F_e'\ ,\\
\overline F_e &=& \cos^2\theta_{12} \overline F'_e+\sin^2\theta_{12}\overline F^0_x\ .
\end{eqnarray}
%................................
Finally, conservation of the total $\nu$ and $\overline\nu$ flux 
provides $F_x$ and $\overline F_x$ by subtraction,
%...............................
\begin{eqnarray}
2F_x &=& (2F_x^0+F_e^0)-F_e\ ,\label {fx}\\
2\overline F_x &=& (2\overline F_x^0+\overline F_e^0)-\overline F_e\label{fxbar}\ .
\end{eqnarray}
%................................

%%%%%%%%%%%%%%%%%%%%%%%%%%% FIGURE 6 %%%%%%%%%%%%%%%%%%%%%%%%%%%%%%%%
\begin{figure}[t]
\centering
\vspace*{4mm}
\hspace*{22mm}
\epsfig{figure=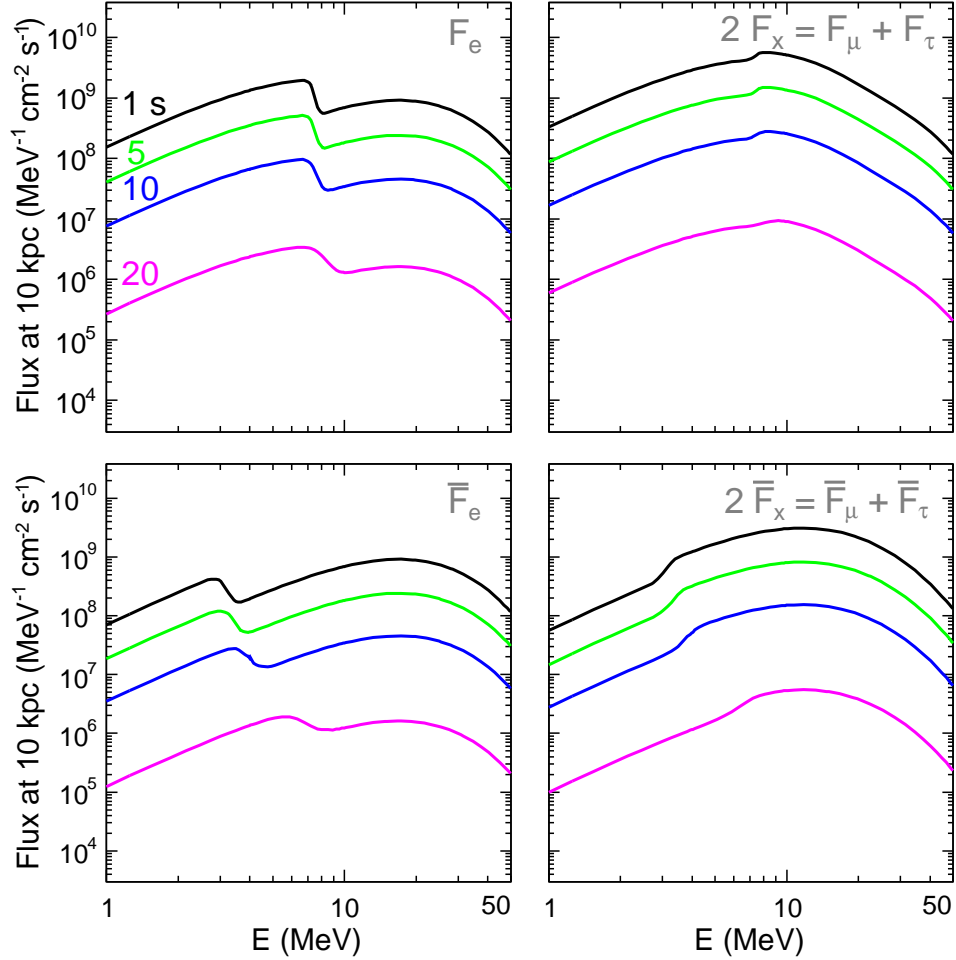,width=0.8\columnwidth}
\vspace*{0mm}
\caption{Final oscillated fluxes of neutrinos ($F_\alpha$) and antineutrinos
($\overline F_\alpha$) at $d=10$~kpc. The comparison with the unoscillated fluxes
in Fig.~1 shows the imprints of collective flavor evolution effects.  
\label{fig 6}}
\vspace*{1cm}
\end{figure}
%%%%%%%%%%%%%%%%%%%%%%%%%%%%%%%%%%%%%%%%%%%%%%%%%%%%%%%%%%%%%%%%%%%%%%

Figure~6 shows the corresponding results  
in graphical form. The spectral split features in the $\nu_e$ and $\overline\nu_e$ spectra 
are somewhat reduced by the low-energy $\theta_{12}$ mixing from
Eqs.~(\ref{lin1}) and (\ref{lin2}), but are still clearly observable at any time $t\geq 1$~s. The $x$-flavor split features 
are instead more suppressed by mixing.

We emphasize that, in view of prospective observations of galactic SN neutrino bursts
(at least for the electron flavor via charged currents)
the persistence of similar stepwise features for several seconds
is rather useful. Detection of such features (if realized in nature) 
requires setting the threshold at low energy ($\lesssim$~few MeV): a very challenging
goal, since one has to fight against backgrounds (which may be large in some
shallow detector projects \cite{5Mega}) and low cross sections.
However, low signal rates may be at least compensated by integration over time 
intervals of a few seconds, 
without canceling the persisting split effects. 
In order to facilitate feasibility studies in specific experimental settings, 
we provide our main results (i.e., the
final fluxes at $d=10$~kpc in Fig.~6, as compared to the unoscillated ones in Fig.~1) 
in computer-readable form upon request. 

\section{Effects of variations in the input SN neutrino spectra}

Our references choice of neutrino energy spectra and luminosities is by no means unique.
Different choices may be motivated by the results of some SN explosion simulations, see e.g. \cite{Janka1}.
We do not consider herein possible deviations from the hypothesis of luminosity equipartition  among 
different neutrino species \cite{Janka1}, which are not supported in our reference time interval ($t=1$-20~s)
by the results of \cite{Totani}, and which
might lead to new and more complicated split features \cite{Dasgupta}
beyond the scope of this paper. However, we do study some interesting variations of the average energies
$\langle E_\alpha \rangle$.

In particular, following Ref.~\cite{Janka2} (inspired by \cite{Janka1}) one may adopt the values:
%..................
\begin{equation}
\langle E_e\rangle=12~\mathrm{MeV},\  
\langle \overline E_{ e}\rangle=15~\mathrm{MeV},\  
\langle E_{x} \rangle=18~\mathrm{MeV} \ (\mathrm{at\ any\ }t)\ ,
\label{refenergies2}
\end{equation}
%.....................
which are much closer to each other than our reference choice in Eq.~(\ref{refenergies}). We 
shall refer to the
choice in Eq.~(\ref{refenergies2}) as to the ``smaller $\Delta \langle E\rangle$ scenario.'' 

One may also include average energies with noticeable changes with time, as suggested, e.g., by
the results in \cite{Totani}, which read:
%..................
\begin{eqnarray}
\langle E_e\rangle=13~\mathrm{MeV},\  
\langle \overline E_{ e}\rangle=16~\mathrm{MeV},\  
\langle E_{x} \rangle=23~\mathrm{MeV} &\ & (t=1~\mathrm{s})\ ;\\
\langle E_e\rangle=11~\mathrm{MeV},\  
\langle \overline E_{ e}\rangle=18~\mathrm{MeV},\  
\langle E_{x} \rangle=25~\mathrm{MeV} &\ & (t=5~\mathrm{s})\ ;\\
\langle E_e\rangle=11~\mathrm{MeV},\  
\langle \overline E_{ e}\rangle=20~\mathrm{MeV},\  
\langle E_{x} \rangle=25~\mathrm{MeV} &\ & (t=10~\mathrm{s})\ ;\\
\langle E_e\rangle=11~\mathrm{MeV},\  
\langle \overline E_{ e}\rangle=20~\mathrm{MeV},\  
\langle E_{x} \rangle=25~\mathrm{MeV} &\ & (t=20~\mathrm{s})\ .
\end{eqnarray}
%.....................
We shall refer to such assignments as to the ``time-dependent $\langle E\rangle$ scenario.''

For fixed energy luminosity [taken as in Eq.~(\ref{lum})], variations in the average neutrino energies
lead to variations in their total number and thus also in the self-interaction potential $\mu$, as compared
to the reference one shown in Fig.~2.
Figure~7 shows the radial profile of $\mu$ for the case of smaller $\Delta \langle E\rangle$ (left panel)
and of time-dependent $\langle E\rangle$ (right panel). The shaded bands corresponds
to the range $\mu\in [\mu_\mathrm{inf},\,\mu_\mathrm{sup}]$ where, according to the
discussion in \cite{Miri}, collective bipolar oscillations occur; the range is time-dependent in
the right panel. Both the $\mu(r)$ curves and the bands in Fig.~7 differ from the corresponding
ones in Fig.~2. In particular, at $t=20$~s, the intersection of the $\mu(r)$ curve with the shaded band in 
Fig.~7 leads to the smallest radial interval ($\Delta r\simeq 15$~km) for the development of bipolar oscillations---a
fact that leads to an interesting consequence, as shown below.

%%%%%%%%%%%%%%%%%%%%%%%%%%% FIGURE 7 %%%%%%%%%%%%%%%%%%%%%%%%%%%%%%%%
\begin{figure}[t]
\centering
\vspace*{4mm}
\hspace*{22mm}
\epsfig{figure=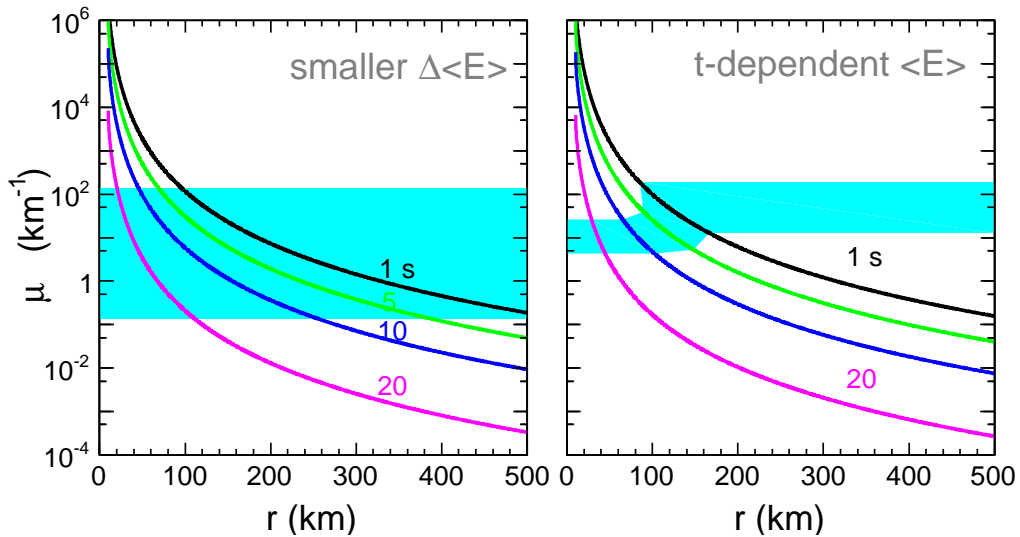,width=0.85\columnwidth}
\vspace*{0mm}
\caption{Radial profiles of the
self-interaction potential $\mu$ at $t=1$, 5, 10, and 20~s for two different variations of the supernova
neutrino input (as compared 
with Fig.~2). Left: case with smaller differences among (constant) average energies. Right: case with
time-dependent average energies. The shaded bands
mark the $\mu$ range where bipolar effects develop. The band acquires a time dependence in the right panel.
See the text for details. 
\label{fig 7}}
\vspace*{1cm}
\end{figure}
%%%%%%%%%%%%%%%%%%%%%%%%%%%%%%%%%%%%%%%%%%%%%%%%%%%%%%%%%%%%%%%%%%%%%%

Figure~8 shows the results of the neutrino flavor evolution (at the end of collective effects) 
in the case of smaller $\Delta \langle E\rangle$. Compared with our reference scenario in Fig.~3, 
the spectral split features are qualitatively similar but less pronounced, as expected 
from the fact the spectral differences of different species are smaller. In particular,
the step-like variation across the split is reduced by a factor of two or more, for
both neutrinos (around $E_c\simeq 7.5$~MeV) and antineutrinos. From the experimental viewpoint, the smaller 
the split variations, the higher the energy resolution and the statistics needed
to observe them. Future SN explosion simulations will shed new light on the 
expected (dis)similarities among the unoscillated spectra of $\nu_e$, $\overline\nu_e$ and $\nu_x$,
and thus on the size of observable oscillation effects (collective or not).

%%%%%%%%%%%%%%%%%%%%%%%%%%% FIGURE 8 %%%%%%%%%%%%%%%%%%%%%%%%%%%%%%%%
\begin{figure}[t]
\centering
\vspace*{4mm}
\hspace*{22mm}
\epsfig{figure=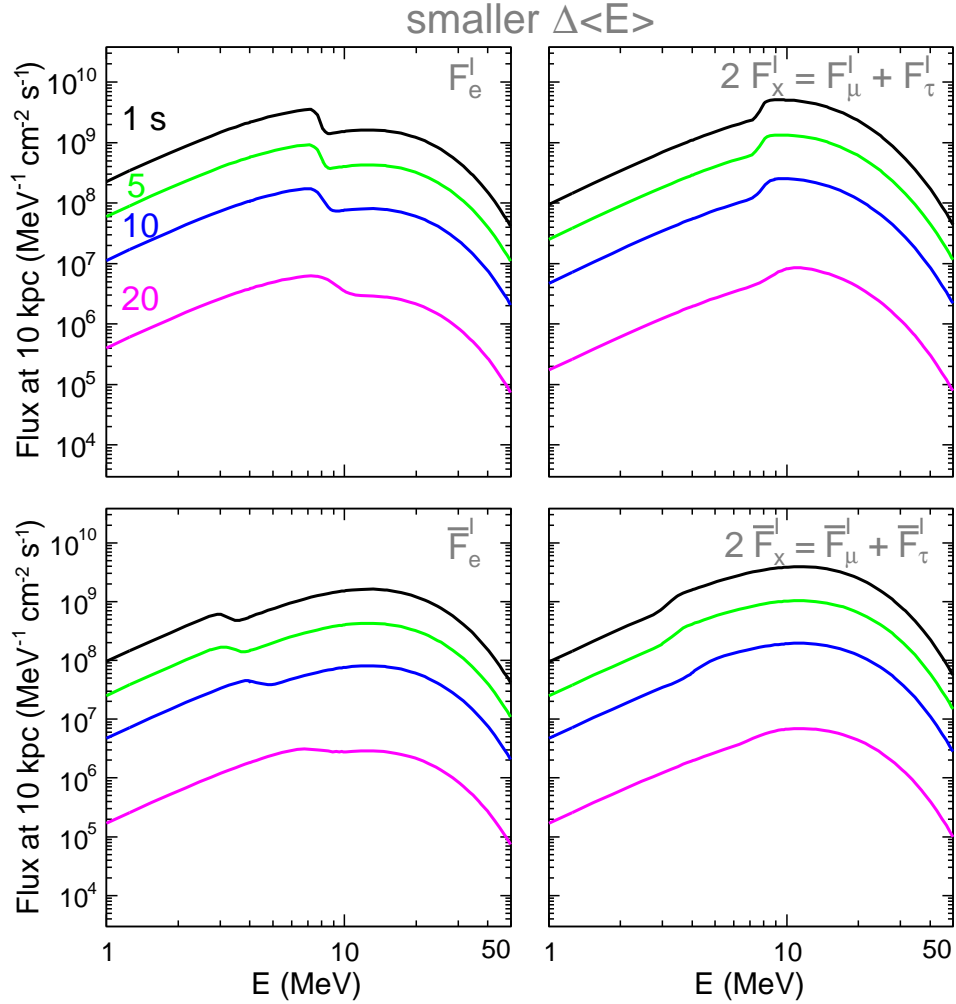,width=0.8\columnwidth}
\vspace*{0mm}
\caption{Fluxes of neutrinos ($F'_\alpha$) and antineutrinos
($\overline F'_\alpha$) at the end of collective effects, 
for the case with smaller $\Delta \langle E\rangle$ (as compared with
the reference scenario in Fig.~3).   
\label{fig 8}}
\vspace*{1cm}
\end{figure}
%%%%%%%%%%%%%%%%%%%%%%%%%%%%%%%%%%%%%%%%%%%%%%%%%%%%%%%%%%%%%%%%%%%%%%

Figure~9 is analogous to Fig.~8, but refers to the case
of time-dependent $\langle E\rangle$ (to be compared with the reference scenario in Fig.~3).
In this case, the critical split energies appear to be also time-dependent. For neutrinos,
conservation of net lepton number (assuming no $\overline\nu$ split in first approximation) 
predicts $E_c\simeq 7.2$, 8.9, 9.7 and 9.7~MeV for the chosen spectra at $t=1$, 5, 10 and 20~s.
This trend is consistent with the numerical results in the upper panels of Fig.~9.
Antineutrinos (lower panels) also show a steadily increasing split energy, which we are unable
to estimate a priori, however. The increase in critical energies with time may favorable from the
experimental viewpoint, since the cross section increase compensate in part the luminosity decrease;
on the other hand, it may prevent integration of spectra on time intervals larger than $\Delta t\sim\! 1$~s, 
where the split feature would be blurred.

Interestingly, no split appear in Fig.~9 at $t=20$~s: the spectra are basically
unoscillated. This fact is related to the very short range $(\Delta r\simeq 16$~km) expected for the
development of bipolar oscillations at $t=20~s$, as noted above. It turns out that the period of bipolar oscillations
is of the same order of $\Delta r$ in this case and, literally, the ``flavor pendulum'' has not enough
time to make a single swing: it remains frozen in the upward, unstable equilibrium position, due
to the extremely sudden decrease of $\mu(r)$. Conversely, in our reference scenario at $t=20$~s (see Fig.~3)
the range $\Delta r$ happens to be slightly larger, thus enabling the flavor pendulum to make a full
swing and to start the collective transitions, which eventually
lead to the spectral split. At late time, the neutrino and antineutrino 
splits appears thus to be relatively fragile phenomena,
whose absence or presence might provide, in principle, some information about 
the gradient of the self-interaction neutrino potential $\mu(r)$.

%%%%%%%%%%%%%%%%%%%%%%%%%%% FIGURE 9 %%%%%%%%%%%%%%%%%%%%%%%%%%%%%%%%
\begin{figure}[t]
\centering
\vspace*{4mm}
\hspace*{22mm}
\epsfig{figure=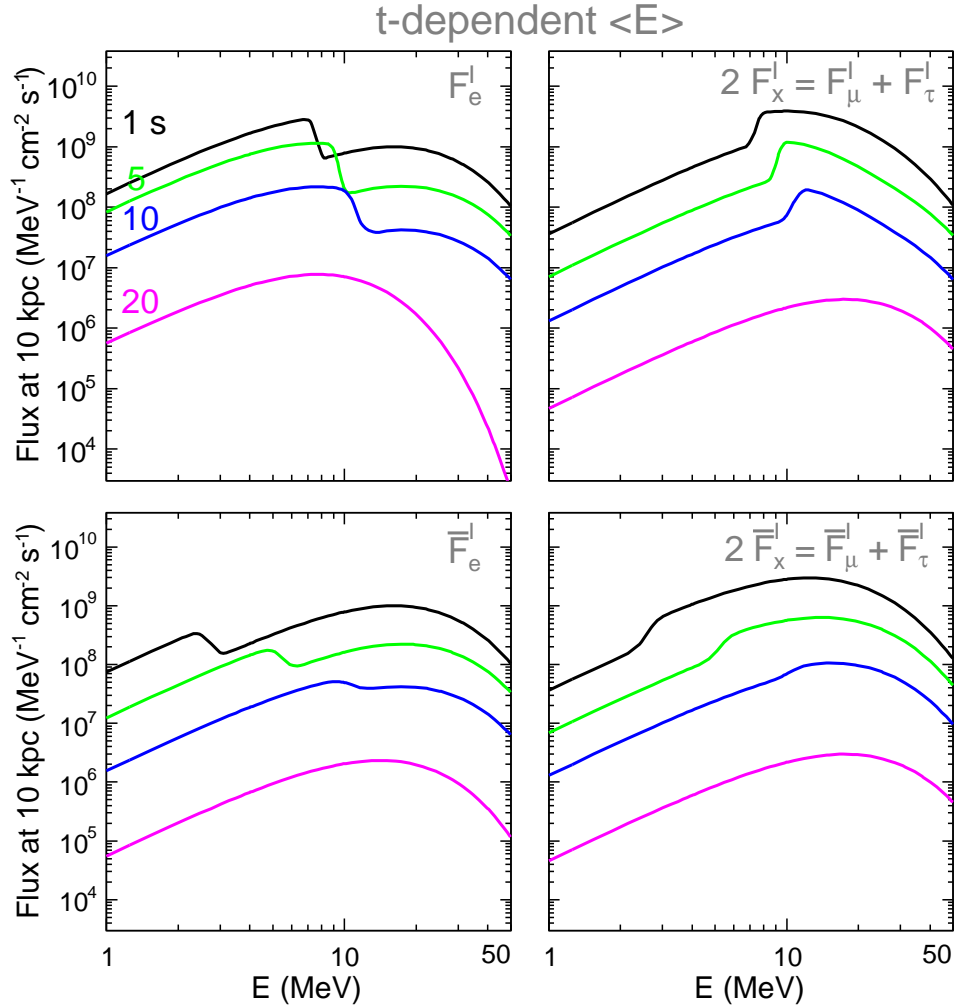,width=0.8\columnwidth}
\vspace*{0mm}
\caption{Fluxes of neutrinos ($F'_\alpha$) and antineutrinos
($\overline F'_\alpha$) at the end of collective effects, 
for the case with time-dependent $\langle E\rangle$ (as compared with
the reference scenario in Fig.~3).  
\label{fig 9}}
\vspace*{1cm}
\end{figure}
%%%%%%%%%%%%%%%%%%%%%%%%%%%%%%%%%%%%%%%%%%%%%%%%%%%%%%%%%%%%%%%%%%%%%%

%%%%%%%%%%%%%%%%%%%%%%%%%%%%%%%%%%%%%%%%%%%%%%%%%%%%%%%%%%%%%%%%%%%%%%%%%
\section{Summary}
%%%%%%%%%%%%%%%%%%%%%%%%%%%%%%%%%%%%%%%%%%%%%%%%%%%%%%%%%%%%%%%%%%%%%%%%

Building upon recent literature on $3\nu$ collective effects in core-collapse supernovae 
\cite{Neutroniz,Esteban,Das3,Full3,Basu,Earth,SNDB,Gava},
we have performed an independent study of three-flavor evolution of neutrinos and antineutrinos 
within a rather ``standard'' SN model (Figs.~1 and 2). Then, assuming inverted neutrino mass 
hierarchy and tiny $\theta_{13}$ (i.e., strongly nonadiabatic $H$-resonance),
self-interaction effects are expected to provide the dominant spectral features, in the form of 
``splits''---or ``stepwise swaps''---for both $\nu$ and $\overline\nu$. 

We have explored this SN $3\nu$ scenario by solving 
the evolution equations for the $3\nu$ density matrix in vector form and single-angle
approximation, after discretization in energy.
The numerical results for the evolved fluxes at the end of collective effects (Fig.~3)
confirm basic expectations at low and high energy [Eqs.~(\ref{je})--(\ref{jxbar})] 
in terms of unevolved
fluxes (Fig.~4). Effects of the ``solar'' mass splitting
and of the $\nu_\mu$-$\nu_\tau$ interaction energy difference in matter are found
to be negligible; effects due to variations of $\theta_{23}$ are large (Fig.~5) but 
unobservable.
The final fluxes at the detector (Fig.~6) are obtained by standard application
of adiabatic $L$-resonance effects.

Both $\nu$ and $\overline\nu$ spectral split features tipically 
emerge with similar characteristics in the whole time interval considered ($t=1$--20~s). 
Observations of such features (if realized in nature) and of their possible variations 
require low-energy thresholds, where the signal is suppressed 
by low cross sections; however, time integration can
partly overcome the suppression, without necessarily 
canceling the persistent spectral features. If the collected statistics 
and the energy resolution allow, spectra collected 
at different times might even reveal migrations
of the split energies due to time variations of neutrino temperatures.
The low-energy frontier in SN neutrino physics may thus be the key to access
the physics of inverted mass hierarchy, especially if $\theta_{13}$ is very small.

\medskip
\medskip
\medskip

%%%%%%%%%%%%%%%%%%%%%%%%%%%%%%%%%%%%%%%%%%%%%%%%%%%%%%%%%%%%%%%%%%%%%%%%%%%%%%%%%
\section*{Acknowledgments}
 %%%%%%%%%%%%%%%%%%%%%%%%%%%%%%%%%%%%%%%%%%%%%%%%%%%%%%%%%%%%%%%%%%%%%%%%%%%%%%%%%%%
This work is supported in part by
the Italian  ``Istituto Nazionale di Fisica Nucleare'' (INFN) and  ``Ministero dell'Istruzione, 
dell'Universit\`a e 
della Ricerca''  (MIUR) through the ``Astroparticle Physics'' research project. I.T.\ acknowledges
support by the E.U.\ (ENTApP network) at the Fifth Annual ENTApP meeting ``Neutrinos in Particle,
in Nuclear, and in Astro-Physics'' (ECT*, Trento, Italy, 2008), where preliminary results of this
work were presented. We thank A.~Mirizzi for careful reading of the manuscript and for 
very useful remarks and suggestions.

%\clearpage

\newpage
%%%%%%%%%%%%%%%%%%%%%%%%%%%%%%%%%%%%%%%%%%%%%%%%%%%%%%%%%%%%%%%%%%%%%%
\section*{References} %%%%%%%%%%%%%%%%%%%%%%%%%%%%%%%%%%%%%%%%%%%%%%%%
%%%%%%%%%%%%%%%%%%%%%%%%%%%%%%%%%%%%%%%%%%%%%%%%%%%%%%%%%%%%%%%%%%%%%%

  \end{document}